\documentstyle[11pt]{article} 

\makeatletter \@addtoreset{equation}{section} \makeatother

\def\dalemb#1#2{{\vbox{\hrule height .#2pt
        \hbox{\vrule width.#2pt height#1pt \kern#1pt
                \vrule width.#2pt}
        \hrule height.#2pt}}}

\def\ben{\begin{equation}}
\def\een{\end{equation}}
\def\half{{\textstyle{1\over2}}}
\let\a=\alpha \let\b=\beta   
  \let\q=\theta  
  \let\n=\nu

\let\C=\Chi 
\let\la=\label  
\def\nn{\nonumber} \def\bd{\begin{document}} \def\ed{\end{document}}
\def\ds{\documentstyle} \let\fr=\frac \let\bl=\bigl \let\br=\bigr
\let\Br=\Bigr \let\Bl=\Bigl
\let\bm=\bibitem
\let\na=\nabla
\let\pa=\partial \let\ov=\overline
\def\ie{{\it i.e.\ }}
\newcommand{\be}{\begin{equation}}
\newcommand{\ee}{\end{equation}}
\def\ba{\begin{array}}
\def\ea{\end{array}}
\def\ft#1#2{{\textstyle{{\scriptstyle #1}\over {\scriptstyle #2}}}}
\def\fft#1#2{{#1 \over #2}}
\def\del{\partial}
\def\sst#1{{\scriptscriptstyle #1}}
\def\oneone{\rlap 1\mkern4mu{\rm l}}
\def\td{\tilde}
\def\wtd{\widetilde}
\def\im{{\rm i}}
\def\bog{Bogomol'nyi\ }
\def\q{{\tilde q}}
\def\hast{{\hat\ast}}
\def\0{{\sst{(0)}}}
\def\1{{\sst{(1)}}}
\def\2{{\sst{(2)}}}
\def\3{{\sst{(3)}}}
\def\4{{\sst{(4)}}}
\def\5{{\sst{(5)}}}
\def\6{{\sst{(6)}}}
\def\7{{\sst{(7)}}}
\def\8{{\sst{(8)}}}
\def\n{{\sst{(n)}}}
\newcommand{\w}[1]{\\[0.#1cm]}
\def\hA{\hat{\cal A}}
\def\ns{{\sst {\rm NS}}}
\def\rr{{\sst {\rm RR}}}
\def\tH{{\widetilde H}}
\def\tB{{\widetilde B}}
\def\cA{{\cal A}}
\def\cF{{\cal F}}
\def\tF{{\wtd F}}
\def\v{{{\cal V}}}
\def\Z{\rlap{\sf Z}\mkern3mu{\sf Z}}
\def\ep{{\epsilon}}
\def\IIA{{\rm IIA}}
\def\IIB{{\rm IIB}}
\def\ads{{\rm AdS}}
\def\R{\rlap{\rm I}\mkern3mu{\rm R}}
\def\ua{\underline{\alpha}}
\def\ub{\underline{\phantom{\alpha}}\!\!\!\beta}
\def\uc{\underline{\phantom{\alpha}}\!\!\!\gamma}
\def\um{\underline{\mu}}
\def\ud{\underline\delta}
\def\ue{\underline\epsilon}
\def\una{\underline a}
\def\unA{\underline A}
\def\unb{\underline b}
\def\unB{\underline B}
\def\unc{\underline c}
\def\unC{\underline C}
\def\und{\underline d}
\def\unD{\underline D}
\def\une{\underline e}
\def\unE{\underline E}
\def\unf{\underline{\phantom{e}}\!\!\!\! f}
\def\unF{\underline F}
\def\ung{\underline g}
\def\unm{\underline m}
\def\unM{\underline M}
\def\unn{\underline n}
\def\unN{\underline N}
\def\unp{\underline{\phantom{a}}\!\!\! p}
\def\unP{\underline P}
\def\unH{\underline{H}}
\def\unF{\underline{F}}
\def\unT{\underline{T}}
\def\ovA{\overline{A}}
\def\ovB{\overline{B}}
\def\uC{{\underline C}}
\def\ns{\normalsize}
\def\vs{\vspace{-0.25cm}}
\def\se{\;\;=\;\;}
\def\de{\;\;:=\;\;}
\def\cF{{\cal F}}
\def\cH{{\cal H}}
\def\cK{{\cal K}}

\def\atwo{\alpha_{2}}
\def\aone{\alpha_{1}}
\def\afive{\alpha_{5}}
\def\ap{\alpha_p}
\def\azero{\alpha_o}
\def\afour{\alpha_{4}}
\def\appt{\alpha_{p+2}}
\def\apmo{\alpha_{p-1}}

\def\cE{{\cal E}}
\def\tr{{\rm tr}}
\def\bC{{\bar \C}}

\newcommand{\bea}{\begin{eqnarray}}
\newcommand{\eea}{\end{eqnarray}}
\newcommand{\ra}{\rightarrow}
\newcommand{\Tr}{{\rm Tr} }
\newcommand{\tamphys}{\it $^\dag$ Michigan Center for Theoretical
Physics\\
Randall Laboratory, Department of Physics, University of
Michigan,\\ Ann Arbor, MI 48109-1120, USA\\
and\\
$^\ddag$Interdisciplinary Center for Theoretical Study\\
University of Science and Technology of China,
 Hefei, Anhui 230026, P.R. China}

\newcommand{\auth}{A.  Batrachenko\footnote{abat@umich.edu}$^\dag$, M.  J.
Duff\footnote{mduff@umich.edu}$^\dag$ and J.  X.  Lu
\footnote{jxlu@umich.edu}$^{\dag \ddag}$}

\begin{document}
\begin{flushright}
\hfill{MCTP-02-nn}\\
\hfill{USTC-ICTS-02-2}\\
\hfill{hep-th/0212186}\\
\end{flushright}
\vspace{10pt}

\begin{center}
{ \large {\bf THE MEMBRANE AT THE END OF THE (DE SITTER)
UNIVERSE\footnote {Research supported in part by DOE grant
DE-FG02-95ER40899.}}.}

\vspace{20pt}

\auth

\vspace{10pt}

{\tamphys}

\vspace{20pt}

\underline{ABSTRACT}

\end{center}

The original {\it membrane at the end of the universe} corresponds to a
probe $M2$-brane of signature $(2,1)$ occupying the $S^{2} \times S^{1}$
boundary of the $(10,1)$ spacetime $AdS_{4} \times S^{7}$, and is
described by an $OSp(4/8)$ SCFT.  However, it was subsequently
generalized to other worldvolume signatures $(s,t)$ and other spacetime
signatures $(S,T)$. An interesting special case is provided by the
$(3,0)$ brane at the end of the de Sitter universe $dS_{4}$ which has
recently featured in the $dS/CFT$ correspondence.  The resulting CFT
contains the one recently proposed as the holographic dual of a
four-dimensional de Sitter cosmology.

Supersymmetry restricts $S,T,s,t$ by requiring that the corresponding
bosonic symmetry $O(s+1,t+1) \times O(S-s,T-t)$ be a subgroup of a
superconformal group.  The case of $dS_{4} \times AdS_{7}$ is `doubly
holographic' and may be regarded as the near horizon geometry of $N_{2}$
$M2$-branes or equivalently, under interchange of conformal and R
symmetry, of $N_{5}$ $M5$-branes, provided $N_{2}=2N_{5}{}^{2}$.
The same correspondence holds in the pp-wave limit of conventional $M$-theory.
{\vfill\leftline{}\vfill 

\pagebreak \setcounter{page}{1}

\newpage

\section{\bf Introduction}
\la{World}

The original {\it membrane at the end of the universe}
\cite{Fifteen,BDPS,BD,NST,BSS,BD2,Sutton,BDPS2,Classical,BSTan}
corresponds to a probe $M2$-brane of signature $(2,1)$ occupying the
$S^{2} \times S^{1}$ boundary of the $(10,1)$ spacetime $AdS_{4}
\times S^{7}$ and is described by an $OSp(4/8)$ superconformal field
theory.  However, it was subsequently generalized \cite{BD2,Classical}
to brane worldvolumes with $s$ space and $t$ time dimensions moving in
a spacetime with $S \geq s$ space and $T \geq t$ time dimensions.  The
brane occupies the boundary of a universe of constant curvature so
that the bosonic symmetry is $O(s+1,t+1)\times O(S-s,T-t)$.
Supersymmetry restricts the values of $s,t,S,T$ to those for which
this bosonic symmetry is a subgroup of a superconformal group, and the
resulting superconformal theories have $(s+t)\leq 6$.  For example, as
discussed in \cite{BD2}, the possible signatures of M-theory are
$(10,1)$, $(9,2)$, $(6,5)$, $(5,6)$, $(2,9)$ and $(1,10)$ and the
possible $M2$-branes have worldvolume signatures $(3,0)$, $(2,1)$,
$(1,2)$ and $(0,3)$.  The corresponding superconformal groups are given in
Table \ref{2branes}.

 In view of the connections \cite{Duffads} between the original membrane at
the end of the universe and the $AdS/CFT$ correspondence
\cite{Maldacena,Gubserklebanovpolyakov,Wittenads}, and in view of the
recent interest in a possible $dS/CFT$ correspondence
\cite{Hulltime,Strominger,Stromingerinf}, it is natural to re-examine
the special case of a $(3,0)$ brane at the end of the de Sitter
universe\footnote{In fact the epithet {\it membrane at the end of the
universe} is even more appropriate in the de Sitter case since the
restaurant in Douglas Adams' book ``Restaurant at the end of the
universe'' \cite{Adams} is located at temporal infinity.} $dS_{4}$.
We show that the resulting CFT contains the one recently proposed as
the holographic dual of a four-dimensional de Sitter cosmology
\cite{Larsen, Halyo}.

\begin{table}
\begin{tabular}{llll}
 $(S,T)$&$(s,t)$& Bosonic Symmetry & Supergroup \\
(10,1) & (2,1) & $O(3,2)\times O(8)$ & $OSp(4/8)$ \\(9,2) & (3,0) &
$O(4,1)\times O(6,2)$ & $OSp^*(4/8)$ \\(9,2) & (1,2) & $O(2,3)\times
O(8)$ & $OSp(4/8)$ \\ (6,5) & (2,1) & $O(3,2)\times O(4,4)$ &
$OSp(4/4,4)$ \\
(6,5) & (0,3) & $O(1,4)\times O(6,2)$ & $OSp^*(4/8)$ \\ (5,6) & (3,0) &
$O(4,1)\times O(2,6)$ & $OSp^*(4/8)$ \\ (5,6) & (1,2) & $O(2,3)\times
O(4,4)$ & $OSp(4/4,4)$ \\ (2,9) & (2,1) & $O(3,2)\times O(8)$ &
$OSp(4/8)$ \\ (2,9) & (0,3) & $O(1,4)\times O(2,6)$ & $OSp^*(4/8)$ \\
(1,10) & (1,2) & $O(2,3)\times O(8)$ & $OSp(4/8)$
\end{tabular}
 \caption{$M2$-branes with world-volume signature $(s,t)$ in spacetime
 signature $(S,T)$.}
\label{2branes}
 \end{table}

In the usual signature all extended objects appear to suffer from
worldvolume ghosts because the kinetic term for the $X^0$ coordinate
enters with the wrong sign.  These are easily removed, however (at
least at the classical level) by the presence of diffeomorphisms on
the worldvolume which allow us to fix a gauge where only
positive-norm states propagate e.g.  the light cone gauge for
strings and its membrane analogues.  Alternatively we may identify
the $d$ worldvolume coordinates $\xi^i$ with $d$ of the $D$
space-time coordinates $X^i$ $(i = 1,2,\cdots d)$ leaving $(D -d)$
coordinates $X^I$ $(I=1..D-d)$ with the right sign for their kinetic
energy \cite{BDPS2}.  Of course, this only works if we have one
worldvolume time coordinate $\tau$ that allows us to choose a
light-cone gauge or else set $\tau = t$.

In the same spirit, we could now require absence of ghosts (or rather
absence of classical instabilities since we are still at the classical
level) for arbitrary signature by requiring that the ``transverse" group
$SO(S - s, T - t)$ which governs physical propagation after
gauge-fixing, be compact.  This requires $T = t$. It may be argued, of
course, that in a world with more than one time dimension, ghosts are
the least of your problems. Moreover, in contrast to strings, unitarity
on the worldvolume does not necessarily imply unitarity in spacetime.
This is because the transverse group no longer coincides with the little
group.  (For example, the $(2,1)$ object in $(10,1)$ spacetime and the
$(1,2)$ object in $(9,2)$ spacetime both have transverse group $SO(8)$,
but the former has little group $SO(9)$ and the latter $ SO(8,1)$.)
Reference \cite{BD2} remained agnostic about the physical significance
of non-compact transverse groups, but simply noted that in the compact
case the list of superconformal groups matches those in Nahm's
classification \cite{Nahm}.

However, an attempt to render these theories respectable has been made
in a series of interesting papers by Hull and collaborators
\cite{Hull:1998br,Hulltime,Hull:1998ym,Hull:1998fh,Hull:1999mr,%
Hull:1999mt,Hullde,Hull:2002cv}, in which it is shown that these theories
with unconventional signatures are related to the usual signature by
means of a timelike T-duality.  Hull denotes the $(10,1)$, $(9,2)$ and
$(6,5)$ signatures as $M$-theory, $M^{*}$-theory and $M'$-theory,
respectively.  Unlike $M$-theory, $M^{*}$ and $M'$ theories can admit de
Sitter vacua invariant under a de Sitter supergroup which does not have
unitary highest weight representations.  This lack of unitarity
representations is reflected in the fact that some of the fields have
kinetic terms with the wrong sign.  However, since these theories are
related to the conventional ones by a timelike T-duality they are no
worse than conventional theories compactified in the time direction and
so perhaps might make sense after all.  Thus Hull's suggestion is both
radical and conservative at the same time.  It is radical in invoking
spacetimes with unusual signature but conservative in saying that the
only such theories we need worry about are those dual to the
conventional theories.

It seems that one is forced to this interpretation if one wants to
combine dS/CFT space with supersymmetry \cite{Hulltime}.  Alternatively,
one can be content to look at non-supersymmetric vacua
\cite{Strominger,Stromingerinf}.  In section \ref{Universe3}, we shall
explore both possibilities, having first reviewed the $AdS$ case in
section \ref{Universe}.

Finally in section \ref{dualities} we note that the case of $dS_{4} \times
AdS_{7}$ is `doubly holographic' and may be regarded as the near
horizon geometry of $N_{2}$ $M2$-branes or equivalently, under
interchange of conformal and R symmetry, of $N_{5}$ $M5$-branes,
provided $N_{2}=2N_{5}{}^{2}$.  The same correspondence holds in the
pp-wave limit of conventional $M$-theory.

\section{A probe (2,1)-brane on the boundary of $AdS_{4}\times S^{7}$}
\la{Universe}

\subsection{The AdS background}
\la{background2}
We begin by reviewing the original membrane at the end of the universe
for which the spacetime is $AdS_4 \times S^7$ and for which the
supermembrane occupies the $S^{1} \times S^{2}$ boundary of the $AdS_4$.
It will be useful to regard the geometry as the near-horizon limit of a
stack of M2 branes.

For M-theory in the usual $(10,1)$ signature, the low-energy effective
field theory is $D=11$ supergravity with bosonic action
\be S_M=
\int d^{11} x \sqrt{-g} \left( R - \frac{1}{48}F_4^2 \right) -{1\over 12}
\int A_3\wedge F_4 \wedge F_4.
\la{mmm}
\ee
where $F_{4}=dA_{3}$.  The M2-brane solution is given by \cite{dust}
\[
ds^2=H^{-2/3} (-dt^2+dx_1^2+dx_2^2) + H^{1/3}(dy_3^2 + \dots +
dy_9^2+dy_{10}^2),
\]
\be
A_{012}=H^{-1},
\ee
where $H(y_3,\dots ,y_{10})$ is a harmonic function in the
transverse space.  For a stack of $N_2$ M2 branes at $y=0$, we take
\be
H=1+\left({L_2\over y}\right)^6,
\la{harmonic}
\ee
where $y^2=\delta_{mn}y^{m}y^{n}$ with $m$ running over the spatial indices in
the transverse space, and $L_2 = (2^5 \pi^2 N_2)^{1/6} l_p$ with $l_p$
being the eleven-dimensional Planck length.  The world-volume has
signature $(2,1)$ and the solution has bosonic symmetry $ISO(2,1)
\times SO(8)$.  It is nonsingular at $y=0$ with near-horizon geometry
$AdS_4 \times S^7$.

\subsection{Unitary but non-supersymmetric}
\la{unitary2}

In anticipation of finding in the next section a unitary but non-supersymmetric
$(3,0)$ brane on the boundary of $dS_{4}$, we begin by forgetting the
$D=11$ supergravity origins of the $(2,1)$ brane and simply look at
its bosonic sector in a background of $AdS_{4}$ with a $4$-form that
follows from the $A_{012}$ given above. We write the $AdS_{4}$ metric as
\be
ds_{AdS_{4}}^{2}=R^2_{AdS_4} \left[ d\rho^{2} + \sinh^{2}\rho
\left(d\theta^{2}+\sin^{2} \theta d\phi^{2}\right)- \cosh^{2} \rho\,
dt^{2} \right]
\la{adsmetric}
\ee
and the 4-form as
\be
F_4=3\,R_{AdS_4}^3\, \cosh\rho\,\sinh^2\rho\, dt\wedge d\rho\wedge
\epsilon_2,
\la{FRansatz2}
\ee where $\epsilon_n$ is the volume form of a unit $n$-sphere and the
$AdS_4$ radius is $R_{AdS_4} = L_2/2$.

Now consider a probe $(2,1)$-brane with a
worldvolume action given by the bosonic sector of the $M2$-brane:
\begin{equation}
S_2=T_2\int d^3\xi\biggl[-{1\over2}\sqrt{-\gamma}\gamma^{ij}\partial_i
x^M\partial_j x^N g_{MN}(x) +{1\over2}\sqrt{-\gamma}
+q{1\over3!}\epsilon^{ijk}\partial_i x^M\partial_j x^N\partial_k x^P
A_{MNP}(x)\biggr]\ ,
\la{membranebose1}
\end{equation}
where $T_2$ is the membrane tension, $\xi^i$ ($i=0,1,2$) are the
worldvolume coordinates, $\gamma_{ij}$ is the worldvolume metric and
$x^M(\xi)$ are the spacetime coordinates $(M=0,1,\ldots,10)$.
Following \cite{Seibergwitten} we allow for a non-extremality parameter
$q$ where the BPS condition corresponds to $q=1$.

The embedding of the membrane in $AdS_{4}$ is given by \be
t=\xi^{0},~~~~~~~\theta=\xi^{1},~~~~~~~\phi=\xi^{2},
\la{braneansatz} \ee so that its worldvolume occupies the $S^{1}
\times S^{2}$ section.  Since we are temporarily ignoring the
$S^{7}$, we focus just on the bosonic radial mode $\rho(\xi)$.  We
substitute (\ref{adsmetric}) and (\ref{FRansatz2}) into
(\ref{membranebose1}) to find \be \label{nambugoto} S_{2}=T_{2}
R^3_{AdS_4} \int_{S^{1} \times S^{2}} d^{3}\xi
\left[-\sqrt{-det(\tilde h_{ij}+
\partial_{i}\rho\partial_{j}\rho)} + q \sinh^{3}\rho \right] \ee
where from (\ref{braneansatz}) and (\ref{adsmetric}) \be \tilde
h_{ij}= \left(
\begin{array}{ccc}
-\cosh^{2}\rho&0&0\\
0&\sinh^{2}\rho&0\\
0&0&\sinh^{2}\rho \sin^{2}\theta
\end{array}
\right). \ee We are interested in the $\rho \rightarrow \infty$ limit, where the brane approaches the boundary of
$AdS_{4}$. Then the action (\ref{nambugoto}) becomes \be S_{2}=T_{2} R^3_{AdS_4} \int _{S^{1} \times S^{2}} d^{3}\xi
\sqrt{-h}\left[- \frac{1}{4}~e^\rho ~h^{ij}\partial_{i} \rho
\partial_{j}\rho + \frac{1}{4} ~(1 - 3 q)~e^\rho - \frac 18 e^{3\rho}(1-q) + {\cal O} (e^{-\rho})
\right] \ee where $h_{ij}$ is the metric on $S^1 \times S^2$,
$ds^2 = - dt^2 + d\theta^2 + sin^2\theta\, d\phi^2$. Making the
change of variable \be \rho = \ln \frac{\varphi^2}{2T_2
R^3_{AdS_4}} + 4 T_2^2 R^6_{AdS_4}
\varphi^{- 4} \la{change2} \ee 
where $\varphi$ is the 3-dimensional scalar field with canonical
dimension $1/2$, and ignoring ${\cal O} (\varphi^{-2})$ terms, we
have \be S_{2}= - \int_{S^{1} \times S^{2}}d^{3}\xi
\sqrt{|h|}\left[\frac{1}{2}h^{ij}\partial_{i}\varphi
\partial_{j}\varphi + \frac{R}{16}\varphi^{2}
+ \frac{(1-q)}{64 T_2^2 R^6_{AdS_4}}\varphi^6\right] \la{boundary0},
\la{sgaction}\ee This is just the bosonic singleton action \cite{BD}
with its scalar mass terms and $g\varphi^{6}/6!$ coupling.  Since
the scalar curvature of $S^{1}\times S^{2}$ equals $2$, we recognize
the correct $R\varphi^{2}$ coefficient for Weyl invariance
\cite{BD}.  The coupling constant is given by $g=45(1-q)/4 T_2^2
R^6_{AdS_4}$. Bearing in mind that $T_{2} = 1/(2\pi)^2 l_p^3$ and
$T_2 R^3_{AdS_4} = (2 N_2)^{1/2}/8\pi$, we have $g = 360\pi^2 (1 -
q)/N_2$. As discussed by Seiberg and Witten, for $q > 1$, the system
is unstable against emission of branes, which is perfectly possible
in a non-supersymmetric theory.

\subsection{Unitary and supersymmetric}
\la{susy2}

When $q=1$, the fermionic completion of the worldvolume action (\ref{membranebose1}) is kappa symmetric. This demands
that the background metric $g_{MN}$ and background 3-form potential $A_{MNP}$ obey the classical field equations of
$D=11$ supergravity.  We shall now consider this case with the full is $AdS_{4} \times S^{7}$ background: \be d
s_{11}{}^2 = R_{AdS_4}^2 \, [\,d \rho^2  + \sinh\rho^2 (\,d\theta^2 + \sin^2\theta\, d\phi^2\,) - \cosh^2\rho \,d
t^2\,] + 4 R_{AdS}^2 d\Omega_7^2\,.  \ee We denote the seven angles parameterizing the $S^7$ as $\theta^m$ with $m = 1,
2, \cdots, 7$ and $d\Omega_7^2 = d\theta^m d\theta^n g_{mn}$.  We can no longer limit ourselves to a single radial
mode, but the generalization to include the $S^7$ modes is straightforward.  To keep things simple, we shall continue
to ignore the fermions, however.  Note that the Wess-Zumino, or volume term remains unchanged.  This implies that the
relation between $\rho$ and $\varphi$ is the same as before.  Note also that the radius for the $S^7$ in the full
metric is twice that of $AdS_4$.  With all the
above, we have, for $\varphi \rightarrow \infty$, 
\be S_{2}ï = - \int_{S^1\times S^2} d^3 \xi \sqrt{|h|} \,
\left[\frac 12 h^{ij} (\,\partial_i \varphi \partial_j \varphi +
\varphi^2
\partial_i \theta^m \partial_j \theta^n g_{mn}\,) +
\frac{1}{8} \varphi^2\right]\,, \ee 
If we define
\be \varphi^a = \varphi \omega^a\,, \qquad (a = 1, 2, \cdots 8)\,,
\ee where $\omega^a$ are the 1-forms parameterizing the unit
$S^{7}$ and $\varphi^a \varphi^a = \varphi^2$, then we have \bea
\label{singletonaction} S_{2}=  - \int_{S^1\times S^2} d^3 \xi
\sqrt{|h|} \left[\frac 12 h^{ij}
\partial_i \varphi^\mu \partial_j
\varphi^\nu \eta_{\mu\nu} + \frac{R}{16}\varphi^a \varphi^b
\delta_{ab} \right] \,, \la{manyma} \eea where $\eta_{\mu\nu} =
\delta_{\mu\nu} = (+, +,\cdots, +)$ and  which has manifest
$SO(3,2)$ conformal symmetry and $SO(8)$ $R$-symmetry. It
 is just the bosonic sector of the $OSp(4/8)$ SCFT \cite{BD}.
 (Incidentally, this rectifies a previous inability to derive the
 correct bosonic mass terms for the membrane at the end of the
 universe, necessary to identify it with the $OSp(4/8)$ supersingleton
 action \cite{BDPS,BDPS2} on $S^{1} \times S^{2}$.  The singleton actions in
 \cite{Kallosh,Fre} were defined on a Minkowski background and so
 required no mass terms.)

\section{A probe $(3,0)$-brane on the boundary of $dS_4 \times AdS_7$}
\la{Universe3}
\subsection{The dS background}
\la{background3}

We will now seek the analogue of the $M2$-brane that occurs in the
$(9,2)$ $M^{*}$ theory, whose field theory limit is a
supergravity theory with bosonic action \cite{Hulltime}
\be
S_{M^{*}}=\int d^{11} x \sqrt{g} \left( R + \frac{1}{48}{F_4^2} \right)
-{1\over 12} \int A_3 \wedge F_4 \wedge F_4.
\la{mstar}
\ee
Note that the sign of the kinetic term of $F_4$ is
opposite\footnote{Interestingly enough, this means that $M^{*}$ theory
avoids the objections that are present in
$M$-theory to implementing Hawking's resolution
\cite{Hawking} of the cosmological constant problem \cite{Cosmo}.}
to that of the action (\ref{mmm}).  As
discussed in \cite{Hull:1998fh}, the sign of the kinetic term is
intimately related with the world-volume signatures that can occur.
For example, if the sign of the kinetic term of $F_4$ were reversed in
(\ref{mstar}) to give a Lagrangian $R - {F_4^2/ 48}+...$ in $9+2$
dimensions, there would be a membrane solution with 2+1 dimensional
world-volume, while the action (\ref{mstar}) with the opposite sign
for the $F_4$ kinetic term has brane solutions with world-volume
signatures $(3,0)$ and $(1,2)$, as we shall see below.  The sign of
the $F_4$ kinetic term in actions (\ref{mmm}), (\ref{mstar}) is
determined by supersymmetry.

This theory has a number of Freund-Rubin-type solutions involving the
de-Sitter-type spaces, including $(d + 1)$-dimensional de Sitter space
$dS_{d + 1}$, $(d + 1)$-dimensional anti-de Sitter space $AdS_{d + 1}$,
the $(d + 1)$-dimensional hyperbolic space $H_{d + 1}$, and the two-time
de Sitter space $AAdS_{d + 1}$ which is a generalized de Sitter space
given by (a connected component of) the coset $SO(d,2)/SO(d-1,2)$, with
signature $(d-1,2)$ and isometry $SO(d,2)$.  The solutions are $dS_4
\times AdS_7$, $AAdS_4 \times S^7$, $AdS_7 \times dS_4$ and $AAdS_7
\times H^4$.

There are two solutions of $M^{*}$-theory analogous to the $M2$-brane: a $(3,0)$-brane and a $(1,2)$-brane. In this
paper we consider the $(3,0)$-brane with Euclidean worldvolume. Its solution is given by
\[
ds^2=H^{-2/3} (dx_1^2+dx_2^2+dx_3^2) + H^{1/3}(-dt^2-dt'^2+dy_4^2 +
\dots + dy_9^2),
\]
\be
 A_{123}=H^{-1},
\la{second}
\ee where $H$ is a harmonic function on the transverse space.

For the $(3,0)$-brane, the null-cone $y^2=t^2+t'^2$ divides the transverse
space into two regions, and there are two distinct brane
solutions, in which the transverse coordinate space is restricted to the
region inside or outside the null cone.  In the region $y^2> t^2 +
t'^2$, a natural choice for the time-dependent harmonic function is
\be
H=1+{L_{2}\over (y^2-t^2-t'^2)^3},
\ee
which gives a real solution
(\ref{second}) for $y^2> t^2+t'^2$.  For $y^2< t^2+t'^2$, we take
instead
\be H=1+{L_{2}\over (t^2+t'^2-y^2)^3}.
\ee
In either case, the
solution has bosonic symmetry $ISO(3) \times SO(6,2)$.  The geometry of
(\ref{second}) near $y^2=t^2+t'^2$ differs in the two cases.

For $t^2+t'^2 > y^2$, let $\tau^2=t^2+t'^2-y^2$. Then near $\tau=0$,
$H^{1/3}\to L_{2}^{1/3}/\tau^2$. Setting $y=\tau \sinh \alpha$, $t=\tau
\cosh\alpha\cos\theta$, and $t'=\tau \cosh\alpha\sin\theta$, the metric
near $\tau=0$ takes the form \be ds^2 ={W^2\over R_{dS_4}^2}
(dx_1^2+dx_2^2+dx_3^2) - {R_{dS_4}^2dW^2\over W^2} + 4 R^2_{dS_4}\left(
d\alpha^2 - \cosh^2\alpha d\theta^2 + \sinh^2\alpha
d\Omega_5^2\right),\ee where $W=L_{2}{}^{-1/6} \tau^2/2$ and again
$R_{dS_4}=L_{2}{}^{1/6}/2 = R_{AdS_7}/2$. This is the metric of $dS_4 \times
AdS_7$. The region $t^2+t'^2 > y^2$ of the solution (\ref{second}) then
interpolates between the flat space $\R^{9,2}$ and $dS_4 \times AdS_7$.
Similar analysis shows that the region $y^2 > t^2 + t'^2$ of the
solution (\ref{second}) interpolates between the flat space $\R^{9,2}$
and $H^4 \times AAdS_7$.

\subsection{Unitary but non-supersymmetric}
\la{unitary3}

 We now wish to repeat the calculation given in the section
 (\ref{unitary2}) but with a $(3,0)$-brane occupying  the $S^{3}$ boundary of
 $dS_{4}$ and with a $4$-form that follows from $A_{123}$ above.  In
 terms of global coordinates, we write the $dS_{4}$ metric and $4$-form
 as 
\bea\label{dSbackground} &&ds^2_{dS_4} = R^2_{dS_4} \,(- d\a^2 + \cosh^2\a\,
d\Omega_3^2\,) \,,\nn\\
&&F_4 = 3 R^3_{dS_4} \cosh^3\a\, d\a\wedge\epsilon_3\,
\la{ds4}
\eea
where $R_{dS_4} = L_2/2$.

Now consider a probe $(3,0)$ brane whose action given by
\begin{equation}\label{action30}
S_2=T_2\int d^3\xi\biggl[-{1\over2}\sqrt{\gamma}\gamma^{ij}\partial_i
x^M\partial_j x^N g_{MN}(x) +{1\over2}\sqrt{\gamma}
+{q\over3!}\epsilon^{ijk}\partial_i x^M\partial_j x^N\partial_k x^P
A_{MNP}(x)\biggr]\ , \la{membranebose3}
\end{equation}
where $i$ now runs over $1,2,3$ and we have once again allowed the
presence of a non-extremality parameter $q$.  The overall sign is
chosen so that the spatial derivatives are the same for $(3,0)$ as for $(2,1)$.
The embedding of the membrane in $dS_{4}$ is given by \ben \xi^1 = \theta^1,\qquad \xi^2 = \theta^2,\qquad \xi^3 =
\theta^3. \een so that its worldvolume occupies the $S^3$ section of $dS_4$ which is located at $\a(\xi)$. We denote
the three angles parameterizing the unit 3-sphere of the $dS_4$ background as $\theta^1, \theta^2$ and $\theta^3$, and
denote its metric by $h_{ij}$.

On the $S^{3}$ boundary of $dS_4$ where $\a$ is large the area term is
\bea T_2
A = \frac{T_2 R^3_{dS_4}}{2^3} \int_{S^{3}} d^3 \xi\sqrt{h} \,\big[e^{3\a} + 3
e^\a - 2 e^\a h^{ij} \partial_i \a \partial_j \a + {\cal O} (e^{-
\a})\big] \eea 
Further the $A_3$ can be solved from $F_4$ given above as
\ben A_3 = R^3_{dS_4} \,(\,3 \sinh \a + \sinh^3\a\,)\, \epsilon_3. \een
so the Wess-Zumino term is 
\bea
&&\ q T_2 \int _{S^{3}} A _3= q T_2 R^3_{dS_4} \int d^3\xi \sqrt{h}(3 \sinh\a +
\sinh^3 \a)\nn\\
&&= \frac{ q T_2 R^3_{dS_4}}{2^3} \int _{S^{3}} d^3\xi \,\sqrt{h}\big[e^{3\a} +
9 e^\a + {\cal O} (e^{- \a})\big].
\eea
Note that since $\alpha$ is a time coordinate, its kinetic term enters
with the opposite sign to that of the spatial coordinate $\rho$ in
section (\ref{unitary2}).  Let us re-express $\a$ in terms of a scalar
field $\varphi$ to eliminate the $e^\a$ in the above Wess-Zumino term.
This can be achieved via 
\ben \a = \ln \frac{\varphi^2}{2T_2 R^3_{dS_4}} - 12 T_2^2
R^6_{dS_4}\varphi^{- 4}\,.  \la{change3} \een 
Note also the different coefficient in the $\varphi^{- 4}$ term relative to that in (\ref{change2}).  This reflects the
fact that $S^{3}$ has a scalar curvature different from $S^{1} \times S^{2}$}.  Ignoring
${\cal O} (\varphi^{- 2})$ terms, the brane action is now 
\bea S_{2} = \int d^3\xi\,\sqrt{h}\left[ \frac 12
(\partial\varphi)^2 + \frac{R}{16} \varphi^2 - (1 -
q)\frac{\varphi^6}{64T_2^2R_{dS_4}^6}\right]\,, \la{emembrane}\eea
Interestingly enough, the above action is also the one (generalized
to a curved boundary) proposed in \cite{Larsen,Halyo} for the
holographic dual of a de Sitter cosmology in the context of a
$dS/CFT$ correspondence \cite{Hulltime,Strominger,Stromingerinf}.
However, the relative sign of the coupling is opposite to that in
the $AdS/CFT$ case of section (\ref{unitary2}) so that the signal
for instability is now $q < 1$.

\subsection{Non-unitary and supersymmetric}
\la{super3}

In closing this section, we consider a probe super $(3,0)$-brane
with $q=1$ on the boundary of $dS_4$ but now in the full $dS_4
\times AdS_7$ background.  So  we consider now other modes of the
brane in addition to the radial one. If we write the metric for
$AdS_7$ with unit radius as \be ds^2_{AdS_7} = d\b^2 - \cosh^2\b\,
d\phi^2 + \sinh^2\b\, d\Omega^2_5 = g_{ij} d \b^i d\b^j \ee
then 
\be
ds_{11}^2 = R^2_{dS_4}\,\big[(-d\a^2 + \cosh^2\a\, d\Omega_3^2) + 4
ds^2_{AdS_7}\big]\,
\la{ds4xasd7}
\ee 
Then we have 
\ben S_{2} = \int_{S^{3}} d^3 \xi \sqrt{h} \left[  \frac 12(\partial \varphi)^2
-\frac 12 g_{ij} \varphi^2
\partial\b^i\partial \b^j + \frac{3}{8} \varphi^2 + {\cal
O} (\varphi^{-2})\right]\,, \een 
where we have used the relation between $\a$ and $\varphi$ given earlier.

The above is not yet the wanted form. For this, we need to define
\bea
&&t = \varphi \cosh \b \sin\phi\,, t' = \varphi \cosh\b \cos\phi\,,\nn\\
&&y^i = \varphi \sinh\b \,\omega^i\,, (i = 1, 2, \cdots 6)\,, \eea 
such that
\ben \varphi^2 = t^2 + t'^2 - y^2\,, \een 
where $y^2 = y^i y^i$.  In the above, $\omega^i$ are the one-forms
parameterizing the unit 5-sphere satisfying $\omega^i \omega^i = 1$.

    One can check explicitly that
\ben dt^2 + dt'^2 - dy^i dy^i = d\varphi^2 - \varphi^2 (d\b^2 - \cosh^2\b\,
d\phi^2 + \sinh^2 \b d\Omega_5^2)\,.  \een 
If we now denote $\varphi^\mu = (t, t', y^i)$, then, ignoring ${\cal O}
(\varphi^{- 2})$ terms, we have the action 
\ben S = - \int_{S^{3}} d^3 \xi\sqrt{h} \left[\frac 12
h^{ab}\partial_a \varphi^\mu \partial_b \varphi^\nu \eta_{\mu\nu}
+ \frac{R} {16} \varphi^\mu\varphi^\nu\eta_{\mu\nu} \right]\,,
\een
where we have also set $\varphi^2 \to - \varphi^2$ and therefore
the signature is now $\eta_{\mu\nu} = (-, -, +, +, +, +, +, +)$.
Once again, the above action has the same form as that given in
(\ref{manyma}). It has manifest $SO(4,1)$ conformal symmetry and
$SO(6,2)$ $R$-symmetry .

\section{Dualities between $M2$ and $M5$-branes in $M$, $M^*$ and $M'$
theories}
\la{dualities}
\subsection{Double holography in $M^{*}$ and $M'$-theory}

An interesting feature of the various $M2$ and $M5$ branes appearing in $M^{*}$ and $M'$ theories is that their near
horizon geometries frequently coincide \cite{Hull:1999mt}.  In the $(9,2)$ spacetime of section (\ref{background3}),
for example, both the $(3,0)$ $M2$-brane and the $(5,1)$ $M5$ brane tend to $dS_{4}\times AdS^{7}$ and share the same
superconformal group $OSp^{*}(4/8)$ whose bosonic symmetry is $O(4,1) \times O(6,2)$.  Note that, in contrast to the
usual $(10,1)$ signature, both factors are conformal groups, and so the system is ``doubly holographic''
\cite{Hull:1999mt}.  Further if we accept the duality between the bulk M-theory and the boundary conformal theory, the
direct consequence of this is that the boundary 3-dimensional conformal theory describes the same physics as the
boundary 6-dimensional conformal theory.  The spacetime symmetry and R-symmetry are exchanged in the two pictures.  As
we shall now demonstrate, however, the requirement that the spacetime may be regarded as the near horizon geometry of
$N_{2}$ $M2$-branes or equivalently, under interchange of conformal and $R$ symmetry, of $N_{5}$ $M5$-branes, imposes
the constraint \be N_{2}=2N_{5}{}^{2} \ee (Although we focus on the concrete example of $(3, 0, -)$-branes and $(5, 1,
-)$-branes in $M^{*}$ theory, the conclusion $N_2 = 2 N_5^2$ is true also for the other examples of double holography.)
The near-horizon geometry for a stack of $N_{2}$ $(3, 0, -)$-branes is given by (\ref{ds4}), whereas the near-horizon
geometry for a stack of $N_{5}$ $(5, 1, -)$-branes is given by \bea &&ds_5^2 = R^2_{AdS_7}\, (d\b^2 - \cosh^2\b\,
d\phi^2 + \sinh^2\b\, d\Omega_5^2\,) +
\frac{R_{AdS_7}^2}{4} \,(- d\a^2 + \cosh\a^2\, d\Omega_3^2\,) \,,\nn\\
&& F_7 = 6 R^6_{AdS_7} \cosh\b\,\sinh^5\b\,d\b\wedge d\phi\wedge
\epsilon_5\,,\nn\\
&&R_{AdS _7} = 2 L_5\,, \eea
where $L_5 = (\pi N_5)^{1/3} l_p$.

To make these two geometries identical, we need to identify the two
metrics and relate the $4$-form $F_4$ and $7$-form $F_7$ via $F_7 = \ast
F_4$ with the $\ast$ denoting the Hodge dual.  This can be achieved
provided 
\ben R_{dS_4} = R_{AdS_7}/2\,, \een
which implies
\ben L_2 = 2 L_5\,, \een
Since $L_2 = (2^5 \pi^2 N_2)^{1/6} l_p$, this gives the relation $N_2
= 2 N_5^2$, as promised.

The above conclusion can be understood as the consequence of the
relation between the two radii, one (denoted as $R_4$) is associated
with 4-dimensional manifold and the other (denoted as $R_7$) with the
7-dimensional one. If the 11-dimensional metric $ds^2 = ds_4^2 + ds_7^2$
describes the near-horizon geometry of either M2 or M5 branes, we always
have $R_4 = R_7/2$. If in a given theory, the M2 branes and the M5
branes have the same near-horizon bulk geometry, it must require $N_2 =
2 N_5^2$ given the relation between $R_4$ and $N_2$ and that between
$R_7$ and $N_5$.

At present, due to our lack of understanding of the theories with more
than one time, it is hard for us to make further checks on this
conjectured $CFT_3/CFT_6$ correspondence.  So the question arises
whether there exists a possibility of such a correspondence, in
certain limit, from the conventional one-time $M$-theory.  The answer
is actually positive thanks to the recent interest in PP-wave
backgrounds.

\subsection{PP waves in M-theory}

Let us recall how the pp-wave limit arises in the conventional $M$-theory.
The near-horizon geometries for $M2$ and $M5$ branes are $AdS_4\times
S^7$ and $AdS_7\times S^4$, respectively.  $AdS/CFT$ correspondence
tells us that $M$-theory on the $AdS_4\times S^7$ background is dual to
a $(2,1)$-dimensional conformal theory while $M$-theory on $AdS_7\times S^4$
is dual to $(5,1)$-dimensional conformal theory.  At first sight, these
look very different.  In global coordinates, the near-horizon geometry for $M2$
branes is 
\bea &&ds_2^2 = R_{AdS_4}^2 \,\left( d\rho^2 + \sinh\rho^2\, d\Omega_2^2
- \cosh\rho^2 d\tau^2\,\right) +
  4 R_{AdS_4}^2 d\Omega_7^2 \,,\nn\\
&& F_4  = -3  R^3_{AdS_4} \cosh\rho\,\sinh^2\rho\,d\rho\wedge
d\tau\wedge\epsilon_2\,,\nn\\
&&R_{AdS _4} =  L_2/2\,, \eea
while the near-horizon geometry for $M5$ branes is 
\bea &&ds_5^2 =  R^2_{AdS_7}\, \left(d\sigma^2  +
\sinh^2\sigma\,d\Omega_5^2 -
\cosh^2\sigma d\tau^2\,\right) + \frac{R^2_{AdS_7}}{4}\,d\Omega_4^2\,,\nn\\
&& F_7 = 6 R^6_{AdS_7} \cosh\sigma\,\sinh^5\sigma\,d\sigma\wedge
d\tau\wedge\epsilon_5\,,\nn\\
&&R_{AdS _7} = 2 L_5\,. \eea
In the above $L_2$ and $L_5$ are the radii of $S^7$ and $S^4$ and are
related to the numbers of $M2$ and $M5$ branes, respectively, as before.

We are now seeking a limit such that the above two geometries have
the same isometries and the unit $2$-sphere in $AdS_4$ of
$AdS_4\times S^7$ can be identified with a unit $2$-sphere in the
$S^4$ of $AdS_7\times S^4$ and the unit $5$-sphere in the $AdS_7$
with a unit $5$-sphere in the $S^7$. This is nothing but the
pp-wave limit.
Following \cite{blau,BMN}, the limit for $AdS_4\times S^7$ is 
\bea &&d s_2^2 = L_2^2 \,\left[2 dx^+ dx^- -
\frac{1}{2}\,\left(\rho^2 + \sigma^2\right)\, (d x^+)^2 +
\frac{1}{4}\left(d\rho^2 + \rho^2\, d\Omega_2^2\right) + d
\sigma^2 +
\sigma^2\,d\Omega_5^2\right] \,,\nn\\
&& \bar F_4 = \frac{3\sqrt{2}}{2^3} L_2^3  \rho^2 d x^+\wedge
d\rho\wedge\epsilon_2 \,, \eea
and the limit for $AdS_7\times S^4$ is 
\bea &&d s_5^2 = 4 L_5^2 \,\left[2 dx^+ dx^- -
\frac{1}{2}\,\left(\rho^2 + \sigma^2\right)\,  (d x^+)^2 +
\frac{1}{4}\left(d\rho^2 + \rho^2\, d\Omega_2^2\right) + d
\sigma^2 +
\sigma^2\,d\Omega_5^2\right] \,,\nn\\
&& \bar F_7 = - 3 2^6 \sqrt{2} L_5^6  \sigma^5 d x^+ \wedge d \sigma
\wedge\epsilon_5 \,. \eea
In the above, the $d\Omega_2^2$ is the original one in $AdS_4$ while the $d\Omega_5^2$ is the one in $AdS_7$. With this
in mind, the above two metrics can be identified provided $L_2 = 2 L_5$. From $\bar F_4 = \star \bar F_7$ we get the
same relation $L_2 = 2 L_5$, which implies $N_2 = 2 N_5^2$. Therefore the conclusion we draw here is that there is a
correspondence between the six-dimensional ${\cal (N_{+},N_{-})}=(2, 0)$ superconformal theory and the
three-dimensional ${\cal N}=8$ superconformal theory, in the respective pp-wave limits of the conventional one-time
M-theory.  The advantage here is that we can make further checks beyond what we can do in the $M^{*}$ case.  By the
respective $AdS/CFT$ correspondence, this must imply that the entropy calculated from the respective bulk theory must
also agree with each other in the pp-wave limit.  For this purpose, we consider both the near-extremal configuration
for black $M2$ branes and that for black $M5$ branes, in terms
of their respective Poincar$\acute{e}$ coordinates \cite{DLP}. For $M2$-branes, 
\bea &&d{\bar s}_2^2 = \frac{4 r^2}{L_2^2} \,\left[ - e^{2f} dt^2 + d
x^i d x^i \right] + \frac{L_2^2}{4} e^{- 2 f} \frac{d
r^2}{r^2} + L_2^2 d\Omega^2_7  \,,\nn\\
&& e^{2 f} = 1 - \frac{r_0^6} {(2 L_2 r)^3} \,, \eea
with $i = 1, 2$, and for $M5$-branes,
\bea &&d{\bar s}_5^2 = \frac{ r^2}{4 L_5^2} \,\left[ - e^{2f} dt^2 + d
x^i d x^i \right] + 4 L_5^2\, e^{- 2 f} \frac{d
r^2}{r^2} + L_5^2 d\Omega^2_4  \,,\nn\\
&& e^{2 f} = 1 - \frac{(4 L_5 r_0)^3} { r^6} \,, \eea
with $i = 1, 2,... 5$.

We now calculate the black hole entropy for the above
configurations, respectively, using the Bekenstein-Hawking entropy
formula $S = A/4 G_{11}$ with $A$ the event-horizon area and
$G_{11}$ the eleven-dimensional Newton's constant\footnote{The
entropy per unit brane volume for  both of the two cases was
calculated in \cite{kles}.}. For $M2$ branes, it is not difficult
to find that the event horizon is located at $r_+ = r_0^2/(2L_2)$
and therefore \bea A &=& \int \frac{4 r_+^2}{L_2^2} L_2^7 d x^1
\wedge d x^2 \wedge
\epsilon_7\,,\nn\\
&=& r_0^4 L_2^3 \int d x^1 \wedge d x^2 \wedge \epsilon_7\,. \eea So we
have the entropy \ben S_2 = \frac{r_0^4 L_2^3}{4 G_{11}} \int d x^1
\wedge d x^2 \wedge \epsilon_7\,. \een For M5 branes, we have from the
above $r_+ = 2 \sqrt{L_5 r_0}$. By the same token, we have the entropy
\ben S_5 = \frac{\sqrt{r_0^5 L_5^3}}{4 G_{11}} \int d x^1\wedge ...
\wedge d x^5\wedge \epsilon_4.  \een In general, it is easy to see that
the entropy $S_2$ is quite different from the entropy $S_5$.

 $AdS_{p + 2}$ can usually be realized as a
hyperboloid in $\R^{2, (p + 1)}$. The metric in terms of the global
coordinates covers the entire hyperboloid while the one in terms of
Poincar$\acute{e}$ coordinates covers only one half of the hyperboloid.
On the one hand, the metric in terms of the Poincar$\acute{e}$
coordinates appears naturally as the near-horizon limit of certain brane
configuration and the corresponding near-extremal thermodynamical
entropy can be calculated while the one in terms of the global
coordinates does not have these properties. On the other hand, the
pp-wave limit is more naturally taken in the $AdS$ metric in terms of
the global coordinates. In order to examine whether there is a
possibility that the two entropies can be set to equal, we need to
relate these two-coordinate descriptions for a given $AdS$-space in the
region for which both descriptions are valid. For a given $AdS_{p + 2}$,
we have \bea r & =& R_{AdS}\, \left(\cosh\alpha\, \sin\tau + \omega_{p +
1}\,\sinh\alpha \right)\,,\nn\\
r\,x^0 &=&R^2_{AdS}\, \cosh\alpha \,\cos\tau\,,\nn\\
r\, x^1 &=& R^2_{AdS} \,\omega_1\,\sinh\alpha \,,\nn\\
\vdots\nn\\
r\, x^p &=& R^2_{AdS}\,\omega_p\, \sinh\alpha \,,\la{crelation} \eea
where $(x^\mu, r)$ with $\mu = 0, 1, ..., p$ are the Poincar$\acute{e}$
coordinates with $(\tau, \alpha, \omega_i)$ are the global coordinates
and  $\sum_i \omega_i^2 = 1$.

For the near-extremal M2 branes, we write $d\Omega_7^2 = d\sigma^2 +
\cos^2\sigma\, d\psi^2 + \sin^2\sigma\, d\Omega_5^2$ and take $\alpha =
\rho$ in the above formulas (\ref{crelation}). Taking the pp-wave limit
($\Omega \to 0$), $\rho \to \Omega \rho$, $\sigma \to \Omega \sigma$ and
$x^- (\equiv (\psi - \tau/2)/\sqrt{2})\to \Omega^2\, x^-$ at $r = r_+$,
we then have \ben dx^1\wedge d x^2 \wedge \epsilon_7 =
\frac{\Omega^9}{2^3} \frac{L_2^8}{r_0^4} \rho^2\sigma^5 d\rho\wedge
d\sigma\wedge \epsilon_2\wedge\epsilon_5\,, \een Then the entropy $S_2$
given above becomes \ben S_2 = \Omega^9 \frac{L_2^9 V_9}{2^3 4
G_{11}}\een which is now independent of the extremal parameter $r_0$ and
where \ben V_9 = \int \rho^2\sigma^5 d\rho\wedge d\sigma \wedge
\epsilon_2\wedge \epsilon_5.\la{volume9}\een  If we define the entropy
in the pp-wave limit as $\bar S_2 = \lim_{\Omega\to 0} S_2/\Omega^9$, then
we have \ben \bar S_2 = \frac{L_2^9 V_9}{2^3 4 G_{11}}.\een

For the near-extremal M5 branes, we write $d\Omega_4^2 = d\rho^2 +
\cos^2\rho\, d\psi^2 + \sin^2\rho\,d\Omega_2^2$ and take now $\alpha =
\sigma$ in (\ref{crelation}).  Taking the pp-wave limit, $\sigma
\to \Omega \sigma$, $\rho \to \Omega \rho$ and $x^- (\equiv (\psi/2 -
\tau)/\sqrt{2})\to \Omega^2\, x^-$ at $r = r_+$, we then have \ben
dx^1\wedge d x^2 ...\wedge d x^5 \wedge \epsilon_4 = \Omega^9
\frac{2^6 L_5^{15/2}}{r_0^{5/2}} \rho^2\sigma^5 d\rho\wedge
d\sigma\wedge \epsilon_2\wedge\epsilon_5\,, \een Then the entropy
$S_5$ becomes \ben S_5 = \Omega^9 \frac{2^6 L_5^9 V_9}{4 G_{11}}\een
which is now also independent of the extremal parameter $r_0$ and
where $V_9$ is also given by (\ref{volume9}).  By the same token, we
have \ben \bar S_5 = \frac{2^6 L_5^9 V_9}{4 G_{11}}.\een
Given our previous discussion about the identification of
       the two metrics under the respective pp-wave limit, we can identify the
       two volume $V_9$ in $\bar S_2$ and $\bar S_5$.  So the two entropies
       can be identified provided $L_2^9/2^3 = 2^6 L_5^9$ which again gives
       $N_2 = 2 N_5^2$.

\section*{Acknowledgements}
We have benefited from conversations with Finn Larsen and Jim Liu.  JXL
acknowledges the partial support of a grant from the Chinese Academy
of Sciences.

\section{Appendix}
\subsection{Large Brane for arbitrary $d$: AdS case}

It is instructive to rewrite the action (\ref{membranebose1}) as \cite{Fifteen}
\begin{equation}
S_2=-T_2\int_{\partial W} d^3\xi\sqrt{|g|}+ qT_{2}\int_{W}F_{4}\ ,
\la{membranebose2}
\end{equation}
where we use the same symbols $g$ and $F_4$ for the metric and 4-form pullbacks onto the
brane.  $W$ is a $4$-manifold whose boundary is the brane
worldvolume $\partial W$. The first term in the action is actually proportional to the area
$A$ of the worldvolume
$\partial M$ of the brane.  Also with $F_{4}$ given by the volume form on $AdS_{4}$
\cite{Duffvan,Aurilia,Freundrubin} in
(\ref{FRansatz2}), the second term can be recognized as proportional to the volume $V$ of $M$
\cite{Fifteen,Seibergwitten}, so that
\begin{equation}
\label{AminusVAdS}
S_{2}=-T_2(A-\frac{3q}{R_{dS_4}}V).
\end{equation}
Following \cite{Seibergwitten}, the results of section
(\ref{unitary2}) may now be generalized to an arbitrary $(d-
1)$-brane occupying the conformal boundary $M_d$ of an arbitrary
Einstein $(d + 1)$-dimensional manifold $W_{d + 1}$ of negative
curvature.

 We shall adapt the notation and the signature used in
\cite{Seibergwitten} for our convenience.  The boundary $M_d$ has a
natural conformal structure but not a natural metric.  Let $h_{ij}$ be
an arbitrary metric on the boundary in its conformal class.  Here the
$\xi^i$, $i=0,\dots,d-1$ are an arbitrary set of local coordinates on
the boundary.  There is then a unique way \cite{graham} to extend the
$\xi^i$ to coordinates on $W_{d + 1}$ near the boundary, adding an
additional coordinate $\rho$ that tends to infinity on the boundary,
such that the metric in a neighborhood of the boundary is \be ds^{2} =
r^2_0\left(d\rho^{2} + \frac{1}{4}e^{2\rho} h_{ij} d\xi^{i} d\xi^{j} -
P_{ij} d\xi^{i} d\xi^{j}+O(e^{-2\rho} )\right) \ee where \be
{P_{ij}={2(d-1)R_{ij}-h_{ij}R\over 2(d-1)(d-2)},} \ee and $R_{ij} $ is
the Ricci tensor of $M_d$, which implies \be {h^{ij}P_{ij}= {R\over
2(d-1)},} \ee

Generalizing (\ref{membranebose2}), a probe $(d - 1)$-brane
occupying the conformal boundary $M (\rho)$ of the $W_{d + 1}$
with the above metric at large $\rho$, the action for a radial
mode can be written as \be S = - T_{d - 1} (A_d - q d V_{d +
1}/r_0) \ee where $T_{d - 1}$ is the brane tension, $A_d$ is the
area of the conformal boundary $M(\rho)$ with respect to the brane
induced metric and $V_{d + 1}$ is the volume of $W_{d + 1}$
enclosed by the conformal boundary.  The area $A_d$ and the volume
$V_{d + 1}$ are all calculated explicitly in \cite{Seibergwitten}.
The action for a radial mode is also given
there: 
\bea
&& S = - T_{d - 1} (A_d - \frac{q d}{r_0} V_{d + 1})\,,\nn\\
&& = \left\{\ba{ll} \frac{T_{d - 1} r_0^d}{2^d} \int d^d \xi
\sqrt{-h}\, \big(-(1-q)\varphi^{\frac{2d}{d-2}} - \frac{8}{(d -
2)^2} [(\partial \varphi)^2 + \frac{(d - 2)}{4 (d - 1)} R
\varphi^2] & \\
 + {\cal O}(\varphi^{\frac{2(d - 4)}{d - 2}})\big)&\rm{for}\quad d > 2 \,,\\
\frac{T_1 r_0^2}{4} \int d^2 \xi \sqrt{-h}\, \left(-(1-q)e^{2\varphi}
- 2 [(\partial \varphi)^2 + \varphi R] + R + {\cal O}
(e^{-2\varphi})\right) & \rm{for}\quad d = 2 \,.\ea \right.  \nn\\
&& \eea
In the above, we have made the change of variable
\ben \rho = \left\{\ba{ll} \frac{2}{d - 2} \ln \varphi +
\frac{1}{(d -
1) (d - 2)} \varphi^{- \frac{4}{d - 2}} R& \rm{for}\quad d > 2\,,\\
\varphi + e^{- 2\varphi} \varphi R & \rm{for}\quad d = 2, \ea
\right.  \la{rpl}\een
where $\varphi$ is the d-dimensional scalar field with canonical
dimension $(d - 2)/2$.

We are interested in $\rho \rightarrow \infty$, i.e., $\varphi
\rightarrow \infty$.  From the above, we find 
\bea
&& S = - T_{d - 1} (A_d - \frac{q d}{r_0} V_{d + 1})\,=\nn\\
&& = \left\{\ba{ll} - \frac{T_{d - 1} r_0^d}{2^d} \int d^d \xi
\sqrt{|h|}\, \left( (1-q)\varphi^{\frac{2d}{d-2}}+ \frac{8}{(d -
2)^2} [(\partial \varphi)^2 + \frac{(d - 2)}{4 (d - 1)} R
\varphi^2]
\right)&\rm{for}\quad d > 2 \,,\\
- \frac{T_1 r_0^2}{4} \int d^2 \xi \sqrt{|h|}\,
\left((1-q)e^{2\varphi} + 2 [(\partial \varphi)^2 + \varphi R] - R
\right) & \rm{for}\quad d = 2
\,,\ea\right.\nn\\  \la{AdS} \eea 
where we recognize the curvature term as that required for Weyl
invariance of the action \cite{BD}. It is also easy to check that
for $d = 3$, we reproduce the singleton action (\ref{boundary0})
derived earlier noting that $r_0 = R_{AdS_4}$.

\subsection{Large Brane for arbitrary $d$: dS case}

Once again we can write the action (\ref{membranebose3}) as
\begin{equation}
S_2=-T_2\int_{\partial W} d^3\xi\sqrt{g}+
qT_{2}\int_{W}F_{4}\ ,
\la{membranebose4}
\end{equation}
so that once again it takes the form of ``area minus volume''
\begin{equation}\label{AminusVdS}
S=-T_2(A-\frac{3q}{R_{dS_4}}V).
\end{equation}

Our probe $(3, 0)$-brane is actually wrapped on the 3-sphere at
a given $\a (\Omega)$.  In the action (\ref{AminusVdS}) the kinetic term can
be essentially viewed as an area integration with the induced metric
while the Wess-Zumino term is proportional to the volume enclosed by
the brane.  Here we check this explicitly and derive the effective
action for the large brane for arbitrary dimension of the de Sitter
factor of the spacetime.  In general, for $dS_{d + 1}$ with metric
\ben ds^2 = r_0^2 \,[- d\a^2 + \cosh^2\a\, d\Omega_d^2\,]\,, \een
we have the area
\bea &&A_d = r_0^d \int d^d \Omega \frac{1}{\sqrt{\det h_{ab}}}
\sqrt{\det(h_{ab} \cosh^2\a - \partial_a \a \partial_b \a)}\,,\nn\\
&& = \frac{r_0^d}{2^d} \int d^d \Omega\, \big[e^{d\a} + d e^{(d - 2)\a} - 2 e^{(d - 2)\a} h^{ab} \partial_a \a
\partial_b \a + {\cal O} (e^{(d - 4)\a})\big]\,, \eea
and the volume
\bea &&V_{d + 1} = r_0^{d + 1} \int d^d \Omega \int_0^\a d \a'
\cosh^d \a' \,,\nn\\
&& = \frac{r_0^{d + 1}}{2^d}\int d^d \Omega \int_0^\a d\a' \big[e^{d\a'}
+ d e^{(d - 2) \a'} + {\cal O} (e^{(d - 4)\a'})\big]  \,,\nn\\
&& = \left\{\ba{ll} \frac{r_0^{d + 1}}{2^d} \int d^d\Omega \big(\frac{1}{d} e^{d\a} + \frac{d}{d - 2} e^{(d - 2)\a} +
{\cal O} (e^{(d -
4)\a})\big) &\rm{for}\quad d > 2\,,\\
\frac{r_0^3}{4}\int d^2 \Omega \big(\half e^{2\a} + 2 \a + {\cal O} (e^{- 2\a})\big) & \rm{for}\quad $d = 2$\ea\,.
\right. \eea
In the above the $r_0$ is the radius of $dS_d$ and $h_{ab}$ is the metric of the unit d-sphere.

If we express the radial mode $\a$ in terms of d-dimensional scalar
field $\varphi$ with canonical dimension $(d - 2)/2$ via 
\ben \a = \left\{\ba{ll} \frac{2}{d - 2} \ln \varphi - \frac{1}{(d -
1) (d - 2)} \varphi^{- \frac{4}{d - 2}} R& \rm{for}\quad d > 2\,,\\
\varphi - e^{- 2\varphi} \varphi R & \rm{for}\quad d = 2, \ea \right.  \een
where $R$ is the curvature of unit d-sphere and $R = d (d - 1)$, we then have
\ben A_d = \left\{\ba{ll} \frac{r_0^d}{2^d} \int d^d x \sqrt{h}\,
\big( \varphi^{\frac{2d}{d - 2}} - \frac{8}{(d - 2)^2} [(\partial
\varphi)^2 + \frac{d - 2}{4(d - 1)} R \varphi^2] + {\cal
O}(\varphi^{\frac{2(d - 4)}{d - 2}})\big)&\rm{for}\quad d > 2\,,\\
\frac{r_0^2}{4} \int d^2 x \sqrt{h} \, \big(e^{2\varphi} - 2
[(\partial \varphi)^2 + R \varphi] + R + {\cal
O}(e^{-2\varphi})\big)&\rm{for} \quad d = 2\,,\ea \right.  \een
and
\ben V_{d + 1} = \left\{\ba{ll} \frac{r_0^{d + 1}}{2^d} \int d^d x
\sqrt{h} \big(\frac{1}{d} \varphi^{\frac{2 d}{d - 2}} + {\cal O}
(\varphi^{\frac{2(d - 4)}{d - 2}}) \big) & \rm{for}\quad d > 2\,,\\
\frac{r_0^2}{4}\int d^2 x \sqrt{h} \big(\half e^{2\varphi} + {\cal O}
( e^{- 2\varphi}) \big) &\rm{for}\quad d = 2\,, \ea \right.  \een 

Finally, the action for the large radial mode $\varphi$ of a probe
$(d, 0)$-brane on the conformal boundary $S^d$ of $dS_{d+1}$
dimensional manifold is given as (ignoring the ${\cal O}
(\varphi^{\frac{2(d - 4)}{d - 2}})$)
\bea
&& S = -T_{d - 1} (A_d - \frac{q d}{r_0} V_{d + 1})\,,\nn\\
&& = \left\{\ba{ll} \frac{T_{d - 1} r_0^d}{2^d} \int d^d x
\sqrt{h}\, (-(1 - q) \varphi^{\frac{2d}{d - 2}} + \frac{8}{(d -
2)^2} [(\partial \varphi)^2 + \frac{(d - 2)}{4 (d - 1)} R
\varphi^2]&\rm{for}\quad d > 2
 \,,\\
\frac{T_1 r_0^2}{4} \int d^2 x \sqrt{h}\, (-(1 - q) e^{2\varphi}
+2 [(\partial \varphi)^2 + \varphi R] - R ) & \rm{for}\quad d = 2
\,.\ea\right.  \nn\\
&& \eea

In analogy with (5.9), we expect this formula to be valid for
arbitrary manifold $W_d$, not just $dS$, but note the change of sign
of the kinetic and mass terms relative to the $AdS$ case. When $d =
3$, we recover (3.13) as expected.

\end{document}